\documentclass[review,12pt,authoryear]{elsarticle}

\usepackage{amssymb}
\usepackage{amsthm}

\usepackage{longtable}
\usepackage{afterpage}
\usepackage{amsmath}
\usepackage{mathtools}
\usepackage{bm}
\usepackage{xcolor}
\usepackage{url}
\usepackage{lscape}

\journal{Computers \& Chemical Engineering}

\begin{document}

\begin{frontmatter}



\title{Convex envelope method for $T,p$ flash calculations for mixtures with an arbitrary number of components and arbitrary aggregate states}

\author[add1]{Quirin Göttl}
\author[add1]{Natalie Rosen}
\author[add1]{Jakob Burger\corref{cor1}}

\address[add1]{Technical University of Munich, Campus Straubing for Biotechnology and Sustainability, Laboratory of Chemical Process Engineering, Uferstr. 53, 94315 Straubing, Germany.}

\cortext[cor1]{Corresponding author e-mail address: burger@tum.de}



\begin{abstract}
$T, p$ flash calculations determine the correct number of phases at phase equilibrium and their compositions for fixed temperature and pressure. They are essential for chemical process simulation and optimization. The convex envelope method (CEM) is an existing approach that employs the tangent plane criterion to determine liquid phase equilibria for mixtures with an arbitrary number of components without providing the number of phases beforehand. This work extends the CEM to include also vapor and solid phases. Thus, any phase equilibrium of a given mixture with an arbitrary number of components and phases can be calculated over the whole composition space. The CEM results are presented for various vapor-liquid and solid-liquid phase equilibria examples of up to four components. We show how the CEM can be used for parameter fitting of $g^E$-models. As an outlook, we demonstrate how the CEM can be combined with a machine learning-based tool for property prediction to construct phase equilibria.
\end{abstract}

\begin{graphicalabstract}
\includegraphics[width=0.9\linewidth]{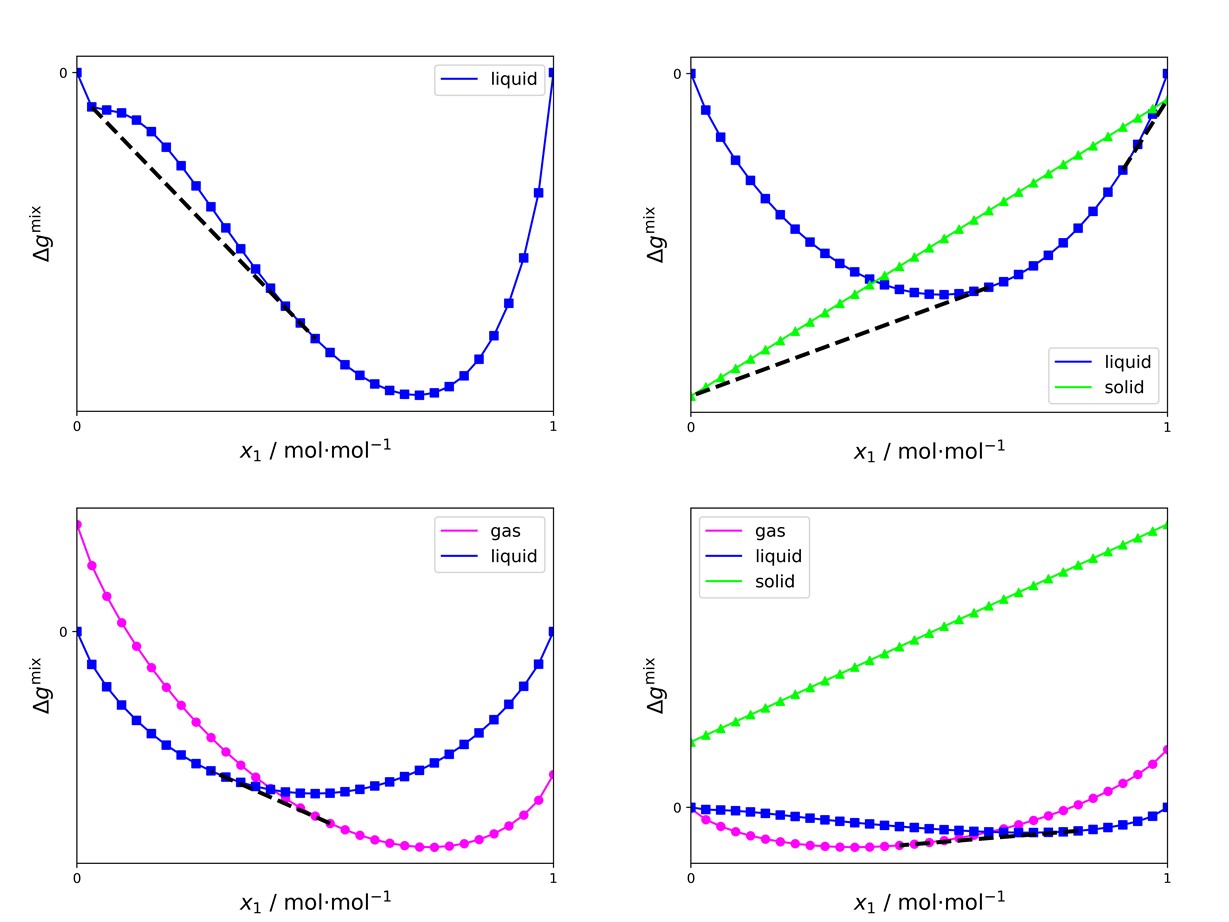}
\end{graphicalabstract}

\begin{highlights}
\item Generalization of the convex envelope method to phase equilibria containing solid, liquid, and gaseous aggregate states.
\item Numerical results for vapor-liquid equilibria and solid-liquid equilibria for mixtures containing up to four components.
\item An open source implementation of the framework in Python.
\end{highlights}


\end{frontmatter}


\theoremstyle{definition}
\newtheorem{defn}{Definition}
\newtheorem{theo}{Theorem}
\newtheorem{rema}{Remark}

\section{Introduction}
\label{introduction}

$T, p$ flash calculations are important in practice to determine all types of phase equilibria \citep{lucia2000}. Calculation of phase equilibria, in turn, is crucial for chemical process simulation and optimization \citep{zhang2011}. 

Generally, the numerical calculation of phase equilibria is an optimization problem. For a $T, p$ flash (i.e., the calculation of the phase equilibrium at constant temperature and pressure) the Gibbs free energy $G$ of the mixture has to be minimized \citep{gibbs1873, gibbs1876}. Thus, a pre-requisite to do a $T, p$ flash is a thermodynamic model of $G$ or other potentials from which it can be derived. There are two main approaches for the respective thermodynamic models: equations of state (EOS) models such as Peng-Robinson \citep{peng1976}, Redlich–Kwong-Soave \citep{soave1972}, SAFT \citep{chapman1989}, or $g^E$ models such NRTL and UNIQUAC \citep{prausnitz1999}. While $g^E$ models are well suited for describing liquid phase behavior, EOS models are applicable to both liquid and vapor phases and provide a continuous transition between them. A detailed discussion of the theoretical foundations of these models is beyond the scope of this work, and interested readers are referred to the sources cited above. We will simply assume that such thermodynamic models are given. The calculations of the $T,p$ flash are greatly simplified, if the types and number of coexisting phases in equilibrium are known. In this case, the solution of the minimization problem can be transformed into an algebraic equation problem, resulting in well known equations such as the (extended) Raoult’s law for the vapor-liquid equilibrium (VLE) or the iso-activity criterion for the liquid-liquid equilibrium (LLE). In the present work, we will consider however the general case, in which the type and number of phases is not known \textit{a priori} and the general optimization problem has to be solved.

We classify the approaches to solve the general optimization problem into two classes. On the one hand, there are mathematical programming approaches which are either deterministic (e.g., \citep{mitsos2007, pereira2010, perevoshchikova2012, wasylkiewicz2013}) or stochastic (e.g., \citep{bonilla-petriciolet2011}). For a more thorough review of such optimization algorithms for phase equilibrium we refer to \citep{zhang2011, piro2016, domanska2019}. 

One the other hand, there are geometrically motivated approaches which find stable phases based on tangents (or more generally convex hulls in case of higher dimensions), on the $G$ state surface (or graph of $G$) in the composition space, resulting in the tangent plane criterion \citep{baker1982, michelsen1982_1, michelsen1982_2}. Examples for such approaches are given in \citep{sorensen1979, swank1986, teh2002}. Another geometrically motivated approach is the class of so-called terrain methods \citep{lucia2002, lucia2003}. These methods are designed to solve chemical process simulation problems, such as nonlinear continuous stirred tank reactors or $T,p$ flash calculations, by intelligently following specific integral curves. In \citep{lucia2005}, terrain methods are applied to a range of multiphase equilibrium problems, including solid–solid equilibria and solid–liquid equilibrium (SLE).

Building on the tangent plane criterion \citep{baker1982}, numerous approaches have been developed based on the convex envelope method (CEM) \citep{ryll2009, ryll2012, goettl2023}. Similar methods are often referred to as the method of the convex hull \citep{lee1992}, which has been applied to the calculation of binary and ternary SLEs. Other works have employed convex hull approaches for the calculation of binary and ternary VLE \citep{hildebrandt1994, afanasyev2012, joseph2017}.

The CEM is fundamentally different from other $T,p$ flash methods. While most of them have the goal to solve the $T,p$ flash problem for a given material feed as efficienctly as possible, the CEM follows a two-step approach. In the first step, which can come with quite some computational effort, the graph of the Gibbs free energy $G$ is computed for the whole discretized composition space of a given mixture. By constructing the convex envelope (or convex hull) for this graph and employing the tangent plane criterion \citep{baker1982}, an approximation (due to the discretization) of the phase equilibrium diagram can be stored in a piecewise linear representation. In the second step, which is in principle decoupled from the first step, $T, p$ flash calculations can be done for any given feed composition extremely fast by solving a small linear problem. Both steps are robust in the sense that they avoid the solution of nonlinear equations or a nonlinear optimization problem. The CEM is a method for determining phase equilibria with an arbitrary number of phases, i.e., the number of phases is determined by the CEM. In \citep{ryll2009, ryll2012}, the CEM was shown to work for liquid phase equilibria for mixtures with up to four components. A graphical evaluation scheme was implemented that distinguishes several cases for the number of existing phases and makes the decision on a case-by-case basis. This work was continued in \citep{goettl2023}, where a general mathematical framework was provided that allows the application of the CEM to mixtures with an arbitrary number of components and an arbitrary number of phases for liquid phase equilibria. The latter two implementations of the CEM were used in computer-aided methods of flowsheet simulation and flowsheet synthesis, relying on the guaranteed robustness of the method and its fast speed once the phase behavior was determined, linearized, and stored for the simulator \citep{ryll2009, ryll2012, goettl2021, goettl2024}. Limitations of the CEM to four components \citep{ryll2009, ryll2012} or only liquid phases \citep{goettl2023} propagated to limitations in the applications, which can be overcome by the present work.

Recently, machine learning (ML) approaches, which treat the underlying thermodynamics as a black box, have gained attention for modeling of phase equilibria. For instance, in \citep{kunde2019}, a support vector machine is employed to develop a surrogate model for LLEs. An artificial neural network is applied in \citep{sun2023} to predict binary VLEs. In \citep{mao2019}, Flory-Huggins theory \citep{flory1941, huggins1941} was combined with elements from the CEM and ML. A convex hull is constructed around the Flory-Huggins free energy density to calculate LLEs for mixtures with up to five components. Here, the simplices within the convex hull are classified using the spectrum of the Laplacian of the underlying adjacency matrices. The authors include a  approach based on principal component analysis and clustering methods to determine the number of coexisting phases and a Cahn-Hilliard approach to model the separation kinetics. For a more comprehensive overview of surrogate models in chemical process engineering and ML approaches for phase equilibrium calculations, we refer to respective reviews \citep{mcbride2019, arroyave2022}.

The present work extends the mathematical framework for the CEM that was provided in \citep{goettl2023}. This allows to simultaneously include all possible phase equilibria between the aggregate states solid, liquid, and gaseous. Given a non-reacting mixture at pressure $p$ and temperature $T$, the framework takes as input the thermodynamic model of $G$ as a function of the composition for all conceivable phases. As output, our framework computes the resulting phase equilibria, i.e., for any given feed composition, it is determined if and which phase equilibrium is present (e.g., VLE, VLLE, LLE, SLE). The number of phases, their aggregate state, and their composition are also calculated.

From a mathematical viewpoint, our method works with mixtures with an arbitrary number of components and an arbitrary number of phases. The CEM is inherently robust, as it relies solely on constructing the convex envelope over a finite set of discretized points spanning the entire composition space. It avoids the need to explicitly enumerate all possible phase equilibria, offering a robust and systematic approach to identifying phase behavior. Once the convex envelope is constructed and the resulting regions are classified, phase splits can be determined reliably and with minimal computational effort. We show the capabilities of our framework with various VLE and SLE examples from the literature with up to four components. We publish the code used to generate all the results shown in this work. Furthermore, we show how the CEM can be utilized to fit $g^E$-model parameters. As an outlook on further usage of the CEM, we combine it with the HANNA-model \citep{specht2024}, a machine learning model that predicts binary activity coefficients from SMILES-strings. Combining the CEM with the HANNA-model, we constructed a complete $T, x, y$-diagram for a binary, heteroazeotropic mixture, ranging from SLE, over LLE, over VLLE, to VLE.

\section{Methodology}

\subsection{General Idea of the CEM}

\begin{figure}
\includegraphics[width=0.9\linewidth]{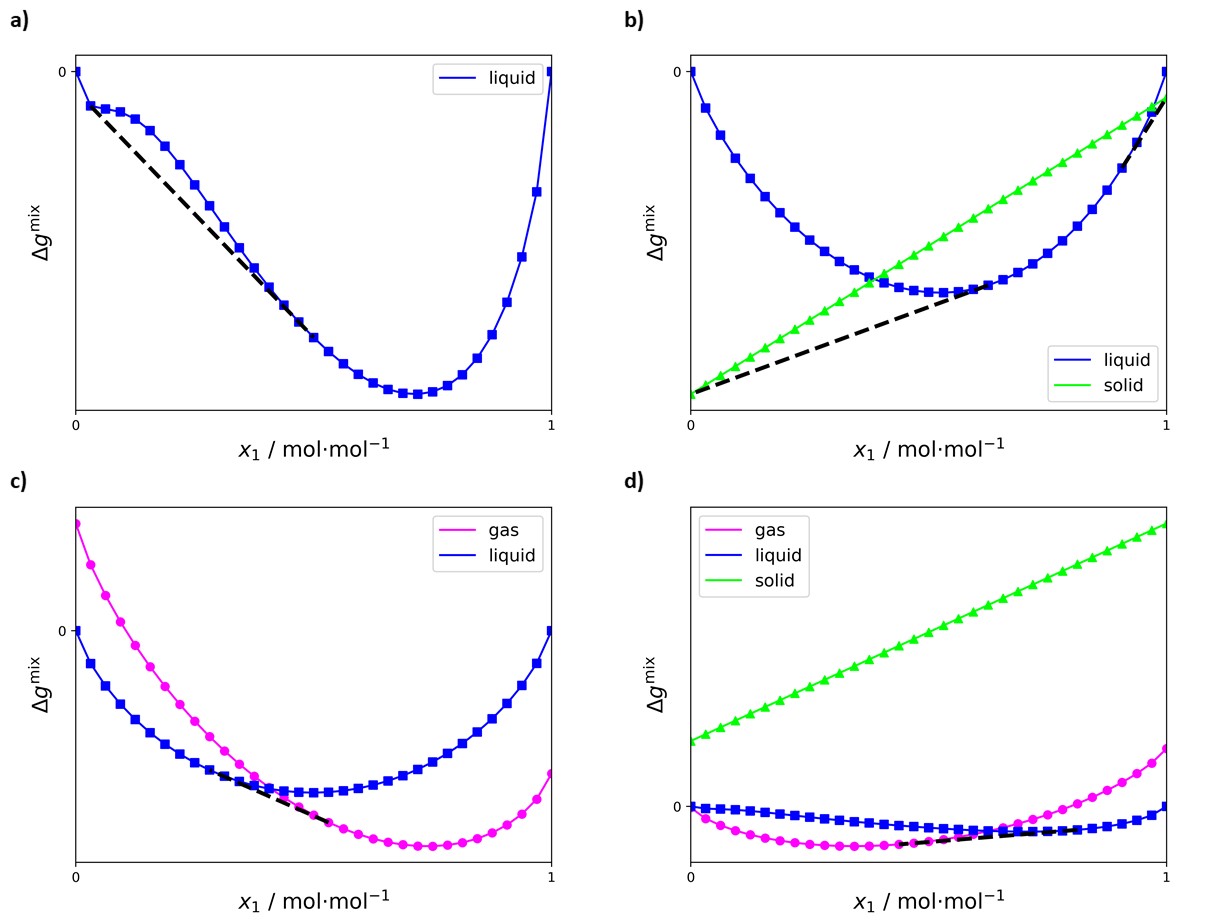}
	\caption{The concept of the convex envelope method is illustrated for an LLE in a), a SLE in b), a VLE in c), and d). The Gibbs energy of mixing is computed for the respective phases, and the convex envelope is constructed for the (combined) graphs to identify phase splits (black dashed line). Blue squares indicate liquid, green triangles solid and pink circles vapor phases. In a), b), and c), the Gibbs energy of mixing graphs of some aggregate states were omitted, since they were not relevant for the construction of the convex envelope (similarly as the Gibbs energy of mixing graph of the solid aggregate state shown in d))}
	\label{figure_cem_concept_for_arbitrary_aggregate_states}
\end{figure}

The idea of the CEM \citep{ryll2009, ryll2012, goettl2023} for liquid phases is illustrated in Figure \ref{figure_cem_concept_for_arbitrary_aggregate_states} a), where the Gibbs energy of mixing for a binary liquid mixture is computed across the entire discretized composition space, represented by blue squares in Figure \ref{figure_cem_concept_for_arbitrary_aggregate_states} a). The convex envelope is then constructed around the graph to identify phase splits. Phase splits occur where the convex envelope connects points that are not direct neighbors in the discretization, as depicted in Figure \ref{figure_cem_concept_for_arbitrary_aggregate_states} a) by the black dashed line (same procedure as described in \citep{goettl2023}). Here, one phase split is observed.
 
To include more aggregate states, the Gibbs energy of mixing for liquid, vapor (pink circles), and solid phases (green triangles) is computed, resulting in one graph for each investigated aggregate state. Convex envelopes are constructed for the combined graphs, as shown in Figure \ref{figure_cem_concept_for_arbitrary_aggregate_states} b), c), and d). For mixtures with multiple aggregate states, phase splits occur in two types. The first type is the same as explained above. In the second type, a phase split is identified when two points connected by the convex envelope belong to different aggregate states, as the aggregate state of a point is determined by the minimum value among the three Gibbs energy of mixing graphs. Examples of this type are shown in Figure \ref{figure_cem_concept_for_arbitrary_aggregate_states} b), c), and d) where one phase split (black dashed line) between distinct aggregate states is identified, respectively. In Figure \ref{figure_cem_concept_for_arbitrary_aggregate_states} a), b), and c), the Gibbs energy of mixing of some aggregate states were omitted for the sake of illustration, since they were not relevant for the construction of the convex envelope (similarly as the graph of the Gibbs energy of mixing of the solid aggregate state shown in Figure \ref{figure_cem_concept_for_arbitrary_aggregate_states} d)).

\subsection{CEM Workflow}

The workflow of the extended CEM is adopted from \citep{ryll2009, ryll2012, goettl2023} and follows the same major steps as introduced in \citep{goettl2023}, with modifications to accommodate the inclusion of vapor and solid aggregate states. In the following, the four steps of the CEM are introduced with a special focus on the modifications. For details of the steps, we refer to \citep{goettl2023}.

\begin{itemize}
	\item [Step I)] \bfseries Discretization of the composition space\mdseries\\
    The discretization is done in the same way as described in \citep{goettl2023}. The composition space of an arbitrary mixture containing $N$ components can be represented by a simplex with $N$ vertices. To discretize this composition space, we specify the distance $1/\delta$, $\delta\in\mathbb{N}$ between neighboring points. Following this procedure for a simple, binary example mixture ($N=2$) with $\delta=2$, we would obtain the following discretization (the elements of the set are the molar fractions of the points in the discretization): $\mathcal{D}=\{(1.0, 0.0), (0.5, 0.5), (0.0, 1.0)\}$. 
	\item [Step II)] \bfseries Determination of the Gibbs free energy graphs and the convex envelope\mdseries\\
    In the present work, we focus on $T, p$ flash calculations and thus the minimization of the Gibbs free energy $G$. Note that for other boundary conditions other potentials (e.g., Helmholtz free energy $A$ for $T, V$ flash calculations) can be used analogously. 

    Our method requires that we can calculate the molar Gibbs free energy $g$ for any composition $\bm{x}$ of any conceivable aggregate phase at given temperature $T$ and pressure $p$: $g=g(T, p, \bm{x})$. For illustration purposes, we chose a commonly used approach based on $g^E$ models. Independent of considered phase $J$, we use pure liquids at $T$ and $p$ as a reference state ("Raoult"):
    
    \begin{equation}
    	g(T, p, \bm{x})={\sum_{i=1}^Nx_ig_i^{\mathrm{pure\,liquid}}(T,p)}+ \Delta g^{\mathrm{mix},J} (T, p, \bm{x}).
    \end{equation}
    
    The sum term of reference states is constant and can thus be omitted for the minimization of Gibbs free energy \citep{ryll2009, ryll2012}. The mixture term $\Delta g^{\mathrm{mix},J}$ contains changes of $g$ related to ideal and non-ideal mixing as well as changes related to phase change if $J$ is not a liquid phase. Under typical assumptions (ideal vapor phase, negligible pressure dependence of non-gaseous properties), one arrives at the following expressions for the liquid and vapor phase \citep{rowlinson1982, olaya2010}:
    \begin{align}
        \label{eq_liquid_dg}
        \Delta g^{\mathrm{mix,liquid}}(T, p, \bm{x})&=\mathcal{R}T\sum_{i=1}^{N}x_i\ln(x_i\gamma_i),\\
        \label{eq_vapor_dg}
        \Delta g^{\mathrm{mix,vapor}}(T, p, \bm{x})&=\mathcal{R}T\sum_{i=1}^{N}x_i(\ln(x_i) + \ln(\frac{p}{p^*_i})).
    \end{align}

    For the solid phase, we assume that the difference between $c_p$ for liquid and solid can be neglected. Additionally, in this work, we only consider mixtures where a pure component can form a solid phase. $\Delta g^{\mathrm{mix,solid}}$ for such a phase consisting only of component $i$ is given by \citep{reyes2001}:
    \begin{equation}
        \label{eq_dg_solid_1}
        \Delta g^{\mathrm{mix,solid}}(T, p)=\frac{T\Delta h_i^f}{T_i^m}(1 - \frac{T_i^m}{T}).
    \end{equation}
    This expression is only defined for pure components, e.g., for $\bm{x}=(1.0, 0.0)$ and $\bm{x}=(0.0, 1.0)$ in a binary mixture. As discussed above, we need an expression for $\Delta g^{\mathrm{mix,solid}}$ that is defined everywhere in the whole composition space. To achieve this, we will use the following expression, which is linear in $x_i$, as can also be seen in Figure \ref{figure_cem_concept_for_arbitrary_aggregate_states} b) and d):
    \begin{equation}
        \label{eq_dg_solid_2}
        \Delta g^{\mathrm{mix,solid}}(T, p, \bm{x})=\mathcal{R}T\sum_{i=1}^{N}x_i\frac{\Delta h_i^f}{\mathcal{R}T_i^m}(1 - \frac{T_i^m}{T}).
    \end{equation}
    The linearity in $x_i$ ensures that the construction of the convex envelope allows solid phases only at the locations of the pure components inside the composition space, e.g., as shown in Figure \ref{figure_cem_concept_for_arbitrary_aggregate_states} b).
    
    In Equations \ref{eq_liquid_dg} - \ref{eq_dg_solid_2}, $\mathcal{R}$ is the ideal gas constant, $\gamma_i$ is the activity coefficient, $T_i^m$ is the melting temperature, $h_i^f$ is the melting heat in J/mol, and $p_i^*$ is the vapor pressure of component $i$. Note that our method also works for more complex thermodynamics, e.g., non-ideal vapor phases or solid complexes as long as $g=g(T, p, \bm{x})$ can be calculated.

    For every point in the discretization $\mathcal{D}$, we compute the Gibbs energy of mixing $\Delta g^{\mathrm{mix}}$ for the liquid, gaseous, and solid aggregate states according to Equations \ref{eq_liquid_dg}, \ref{eq_vapor_dg}, and \ref{eq_dg_solid_2}.

    Combining these three graphs yields a plot as for example shown in Figure \ref{figure_cem_concept_for_arbitrary_aggregate_states} d).
    The molar fractions $\bm{x}$ are transformed to cartesian coordinates $\bm{a}\in\mathbb{R}^n$, $n=N-1$ (for details on this transformation, we refer to \citep{goettl2023}). The transformed molar fractions are combined with the obtained values for $\Delta g^{\mathrm{mix}}$ to obtain the following graph: 
    \begin{equation*}
        \mathcal{G}=\{(\bm{a}, \Delta g^{\mathrm{mix}, s}(\bm{x}))|\bm{x}\in\mathcal{D}, s\in\{\text{solid, liquid, vapor}\}\}\subseteq\mathbb{R}^{n+1}.
    \end{equation*}
    Subsequently, the convex envelope is constructed around the elements of the graph $\mathcal{G}$.
    \item [Step III)] \bfseries Classification of the simplices of the convex envelope\mdseries\\ 
    The convex envelope constructed in Step II) consists of several simplices, each corresponding to structures of different dimensions depending on the number of components in the mixture. In binary mixtures, for example, these simplices are straight line segments; for ternary mixtures, they are triangles; for quaternary mixtures, they are tetrahedra; and so on. Each simplex within the convex envelope must be examined to determine whether it represents a phase split by a linearized model. 
    
    As described in \citep{goettl2023}, this examination consists of two distinct checks for each simplex. The first check involves determining whether the simplex contains at least one heterogeneous line segment. A heterogeneous line segment is defined as a straight line connecting two points with different phases. In \citep{goettl2023}, a line segment was classified as heterogeneous if it connects two points that are not neighboring inside the discretization, e.g., as visualized in Figure \ref{figure_cem_concept_for_arbitrary_aggregate_states} a): the black dashed line is a heterogeneous line segment (and also a simplex of the convex envelope since $N=2$ for this example), since it is part of the convex envelope and connects two points that are not neighboring inside the discretization. In this work, we additionally classify also line segments as heterogeneous that connect points with different aggregate states, e.g., the black dashed line in Figure \ref{figure_cem_concept_for_arbitrary_aggregate_states} c) that connects a point from the liquid and a point from the vapor phase. If the simplex contains at least one such a line segment, it is called a heterogeneous simplex and the second check of the procedure is initiated. 
    
    In the second check, it is evaluated if the simplex models a unique phase split in a linearized phase separator model. As mentioned in \citep{ryll2009, goettl2023}, simplices that fail the second check occur rarely and only at the boundary of the multiphase regions, e.g., close to the critical point. Usually, they nearly vanish, when enough points within the discretization are specified. The second check follows the exact same procedure as outlined in \citep{goettl2023} and is therefore omitted here.

    The simplices that fulfill both criteria are called feasible, heterogeneous simplices.
\end{itemize}

Steps I)-III) have to be executed to yield the phase equilibrium for the given mixture at specified $T$ and $p$. To store the phase equilibrium, it is sufficient to store the feasible, heterogeneous simplices that model a phase split found in Step III). 

\begin{itemize}
	\item [Step IV)] \bfseries Computation of phase splits\mdseries\\  
    For any given feed composition, one has to check if the feed lies within one of the feasible, heterogeneous simplices that model a phase split (from Step III)).
	If such a simplex is found, the resulting phase split can be calculated in the same way as described in \citep{goettl2023}.
\end{itemize}

\subsection{$g^E$-model parameter fitting using the CEM}
\label{section_method_ge_fit}

This section describes how the CEM can be used for fitting parameters of a $g^E$-model to given experimental data. The method is explained alongside the NRTL model \citep{prausnitz1999}, but can also be applied to other models.

According to the NRTL model, the activity coefficient $\gamma_i$ of component $i$ is defined as:
\begin{equation}
	\label{NRTL_equation}
	\ln\gamma_i=\frac{\sum_{j=1}^N\tau_{ji}G_{ji}x_j}{\sum_{l=1}^NG_{li}x_l} + \sum_{j=1}^{N}\frac{x_jG_{ij}}{\sum_{l=1}^{N}G_{lj}x_l}\left(\tau_{ij}-\frac{\sum_{r=1}^{N}x_r\tau_{rj}G_{rj}}{\sum_{l=1}^{N}G_{lj}x_l}\right).
\end{equation}
In Equation \ref{NRTL_equation}, the interactions between components $i$ and $j$ are defined by (we assume for the non-randomness parameter $\alpha_{ij}=\alpha_{ji}=\alpha$):
\begin{align}
\label{NRTL_para_def}
	\tau_{ij}=& \frac{A_{ij}}{RT},\\
	G_{ij}=& \exp(-\alpha\tau_{ij}).
\end{align}

We consider a binary mixture and a set of experimental data points $\mathcal{E}$ that describe the respective phase equilibrium. An optimization procedure that searches for parameters $(A_{12}, A_{21}, \alpha)$ is defined with respect to the following criteria (in case of a mixture with $N$ components, one searches $(A_{ij}, A_{ji}, \alpha)$ for every binary subsystem $(i, j)$):
\begin{itemize}
	\item [1.] For every data point in $\mathcal{E}$ (i.e., $p$, $T$, molar fractions of the feed, molar fractions and aggregate states of the respective phases at equilibrium), the CEM calculates the correct number of phases and the respective correct aggregate states.
	\item [2.] The mean deviation (MD) of the calculated molar fractions for all data points $\mathcal{T}\subseteq\mathcal{E}$ that fulfill the first criterion is minimal:
	\begin{equation}
        \label{objective_NRTL_fit_equation}
		\min_{(A_{12}, A_{21}, \alpha)}\sum_{t\in\mathcal{T}}\sum_{\text{phases } p \text{ in } t}||\bm{x}^{\text{exp}}_p-\bm{x}^{\text{CEM}}_p||_1.
	\end{equation}
	Herein, $||.||_1$ is the $l^1$-norm (i.e., the sum of absolute values of the vector entries).
\end{itemize}

One possible approach to handle those criteria is to interpret the first criterion as a discrete constraint (it is either fulfilled or not fulfilled) and to apply some gradient-based optimization approach to find the minimum in Equation \ref{objective_NRTL_fit_equation}. Since such approaches usually require an initial solution that already fulfills the constraint (which we do not have at hand), we will define an alternative optimization approach as described in the following.

We will utilize a genetic algorithm based on differential evolution \citep{das2011} that works as follows (for the details, we refer to our implementation and to \citep{das2011}):

\begin{itemize}
    \item [1.] A random population of solutions is initialized (every solution is of the form $(A_{12}, A_{21}, \alpha)$).
    \item [2.] For a fixed number of generations, the current population is used to generate new candidate solutions using differential evolution \citep{das2011}. The candidate solutions are compared to randomly chosen members of the population and replace those if they are superior.
    \item [3.] The best solution is returned.
\end{itemize}

This procedure only requires a comparison-function as input that decides for two given solutions which solution is superior (for the second step, when the candidate solutions are pitched against the population members). We define this function as follows for two given solutions $S_1 = (A_{12}^1, A_{21}^1, \alpha^1)$ and $S_2 = (A_{12}^2, A_{21}^2, \alpha^2)$ as a 2-step comparison with respect to the previously defined criteria:

\begin{itemize}
    \item [1.] For both solutions, $S_1$ and $S_2$, it is compared for how many points contained in $\mathcal{E}$ the CEM calculates the correct number of phases and the respective correct aggregate states. If one solution is superior to the other with respect to that criterion, this solution is rated as the better solution and returned. If for both solutions the CEM calculates the correct number of phases and the respective correct aggregate states for the same number of points in $\mathcal{E}$, the solutions are also compared regarding the second criterion.
    \item [2.] Both solutions are compared with respect to Equation \ref{objective_NRTL_fit_equation}. The solution with the lower value is returned as the better solution.
\end{itemize}

Using this comparison function in combination with the genetic algorithm based on differential evolution \citep{das2011}, we can find solutions that provide a good prediction score for most of the available experimental data points. As will be shown in Section \ref{section_ge_fit_results}, the found parameter sets yield similar results as the provided parameter sets from the literature.

\subsection{Implementation}

The whole framework was implemented with the programming language Python (version 3.11.7). The convex envelope is found by usage of the QuickHull-algorithm \citep{barber1996}, which is provided by the Qhull library within the package scipy \citep{virtanen2020}. An implementation of the CEM for arbitrary aggregate states is published via GitHub: \url{https://github.com/quiringoettl/cem4all}.

\section{Results}

\subsection{Qualitative Evaluation}

It has already been shown that the CEM can be used for $T, p$ flash calculations for LLEs \citep{ryll2009, ryll2012, goettl2023}. In the present work, we show that the CEM also works for VLEs and SLEs. For a qualitative evaluation of VLEs and SLEs, the phase equilibria of respective binary and ternary mixtures are constructed using parameters provided in the literature and are qualitatively compared to the source diagrams.

\begin{figure}
\includegraphics[width=0.9\linewidth]{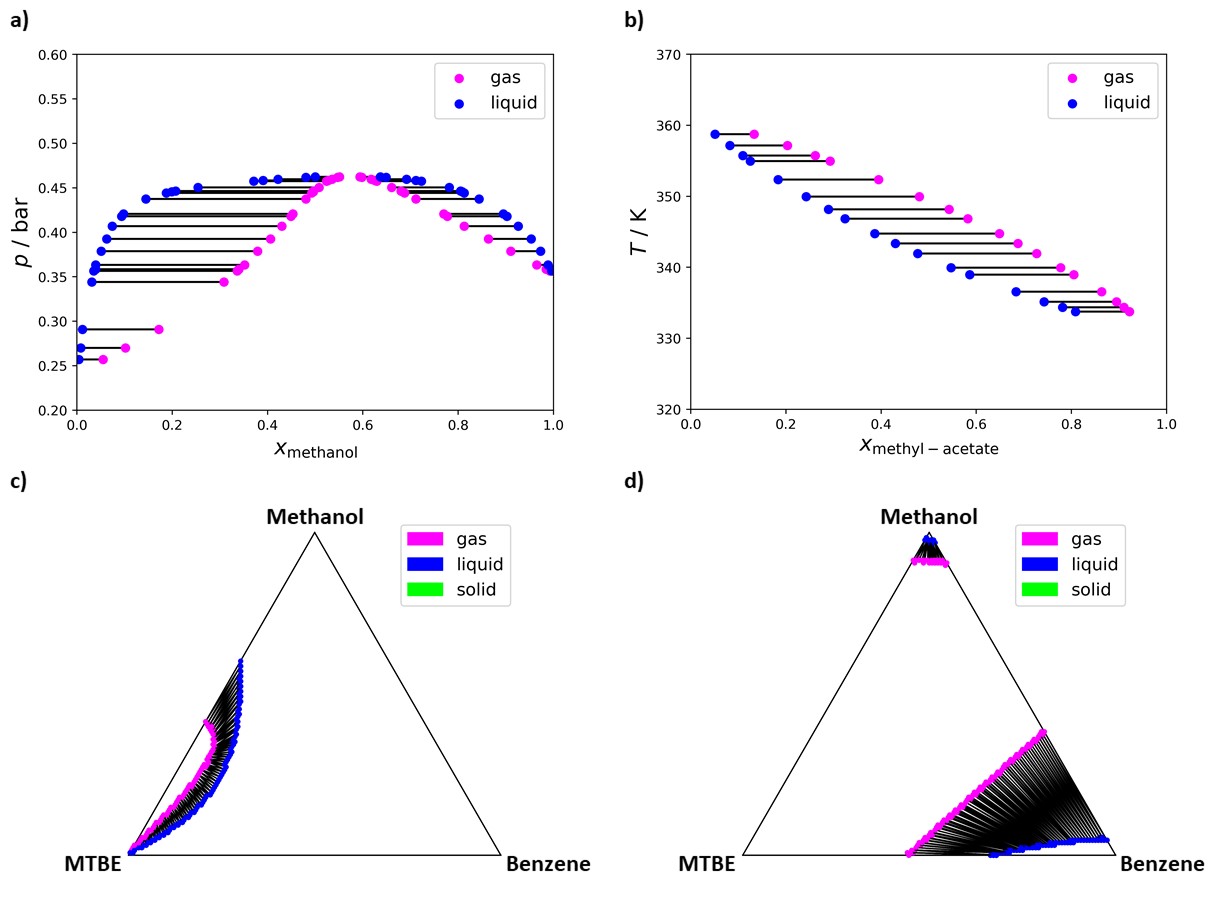}
\caption{Binary and ternary VLEs of a) methanol -- benzene (313.15 K) b) methyl acetate -- isopropyl acetate (1.013 bar), and methyl \textit MTBE -- methanol -- benzene (313.50 K) c) 0.6108 bar d) 0.3854 bar.}
\label{VLE_bin_tern}
\end{figure}

For VLEs, the CEM takes NRTL parameters and Antoine parameters for the pure components as input. In Figure \ref{VLE_bin_tern} a), we show a $p, x, y$-diagram for the binary mixture methanol -- benzene at 313.15 K \citep{oh2003}. Figure \ref{VLE_bin_tern} b) shows the $T, x, y$-diagram for methyl acetate -- isopropyl acetate at 1.013 bar \citep{xiao2013}. Figure \ref{VLE_bin_tern} c) and d) show the VLE of the ternary mixture methyl \textit{tert}-butyl ether (MTBE) -- methanol -- benzene at 313.50 K and 0.6108 bar (Figure \ref{VLE_bin_tern} c)) and 0.3854 bar (Figure \ref{VLE_bin_tern} d)). All plots in Figure \ref{VLE_bin_tern} are generated using the CEM and show the expected phase split behavior.

\begin{figure}
\includegraphics[width=0.9\linewidth]{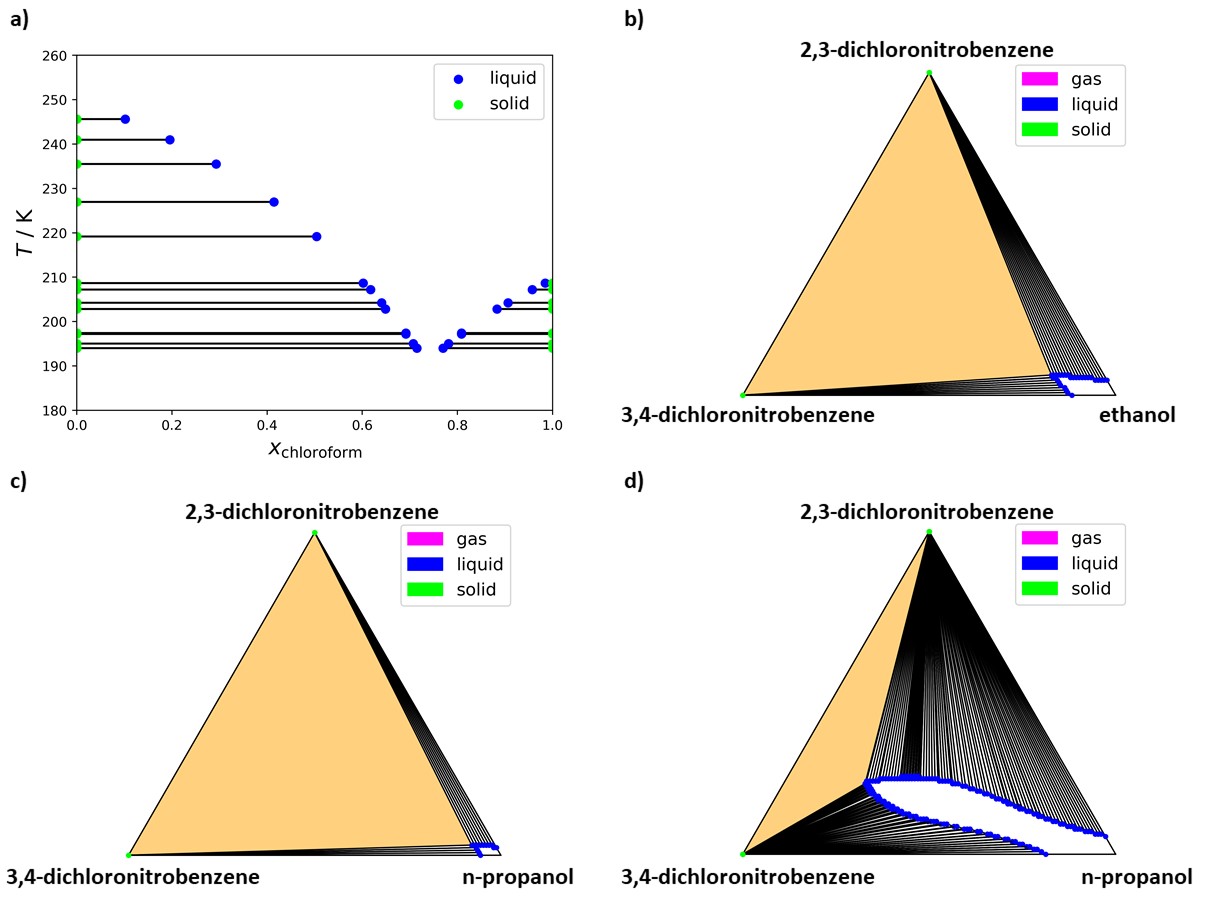}
\caption{Binary and ternary SLEs of a) chloroform -- acetylacetone (1.0133 bar), b) 3,4-dichloronitrobenzene -- 2,3-dichloronitrobenzene -- ethanol (303.15 K and 1.0133 bar) and 3,4-dichloronitrobenzene -- 2,3-dichloronitrobenzene -- n-propanol (1.0133 bar) c) 283.15 K d) 303.15 K. The yellow area marks a split into three phases.}
\label{SLE_bin_tern}
\end{figure}

For SLEs, the CEM takes NRTL parameters, melting temperatures, and melting heats for the pure components as input. In Figure \ref{SLE_bin_tern} a), a $T, x, y$-diagram for the binary mixture chloroform -- acetylacetone at 1.0133 bar \citep{kim2011} is shown. Figure \ref{SLE_bin_tern} b) depicts the SLE of the ternary mixture 3,4-dichloronitrobenzene -- 2,3-dichloronitrobenzene -- ethanol at 303.15 K and 1.0133 bar \citep{li2016}. The yellow area marks a split into three phases: two (pure) solid phases and one liquid phase. In Figure \ref{SLE_bin_tern} c) and d), the SLE of the mixture 3,4-dichloronitrobenzene -- 2,3-dichloronitrobenzene -- n-propanol at 1.0133 bar and at temperatures of 283.15 K in is given in c) and at 303.15 K in d). Once again, the yellow area marks a split into three phases. When comparing those plots to the literature \citep{kim2011, li2016} one can see that the CEM is able to calculate the respective SLEs qualitatively.

\begin{figure}
\label{figure_HANNA}
\includegraphics[width=0.9\linewidth]{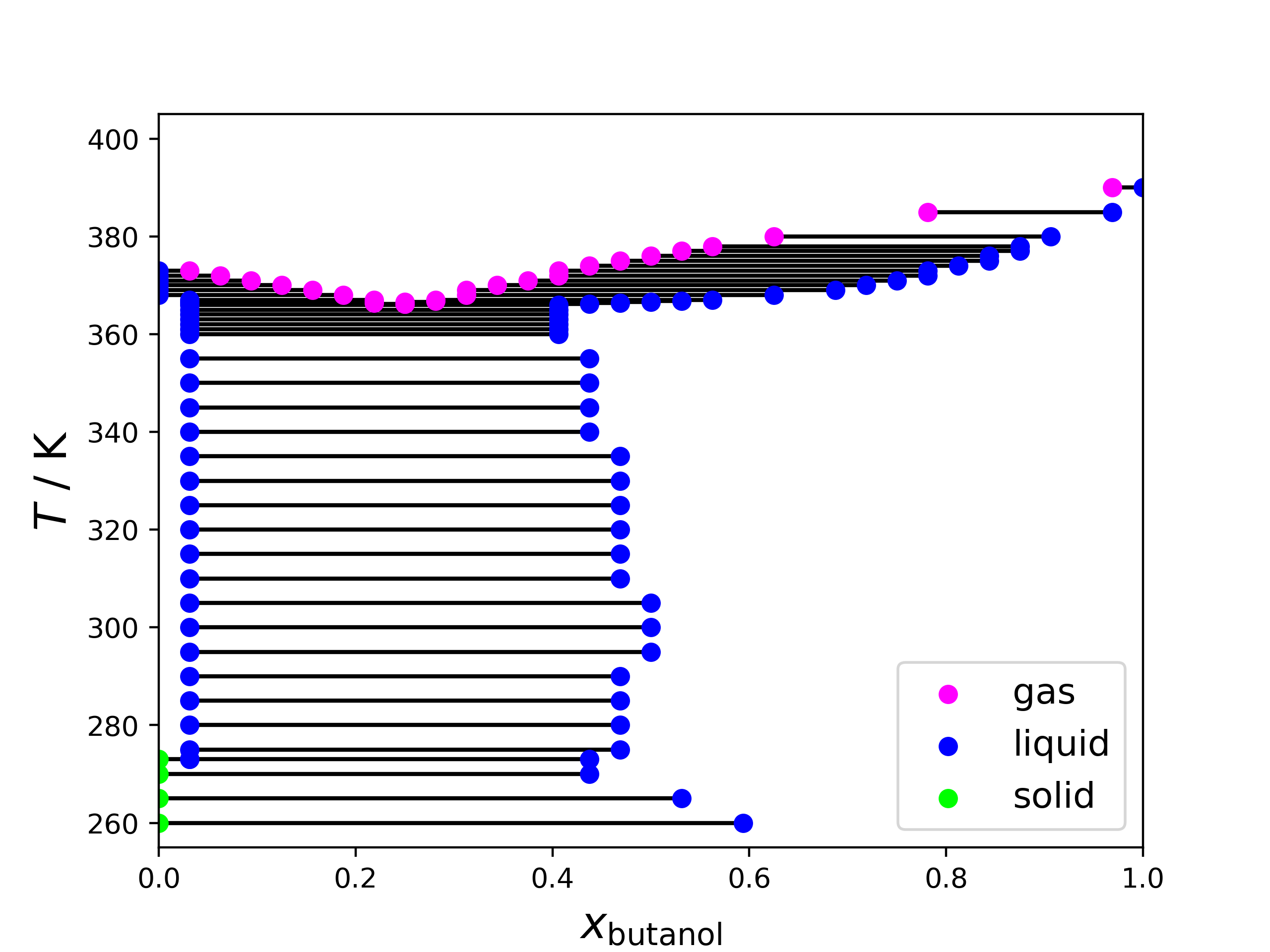}
\caption{$T, x, y$-diagram for the binary mixture butanol -- water ranging from SLE, over LLE, over VLLE, to VLE. The activity coefficients were predicted using the HANNA-model \citep{specht2024}. Antoine parameters, melting temperatures, and melting heats were taken from \citep{nist2024}.}
\end{figure}

When no NRTL model parameters (and also no other $g^E$-model parameters) are available, one can combine the CEM with models that directly predict the activity coefficients for a mixture given its components' SMILES-strings, such as the HANNA-model \citep{specht2024}. In Figure \ref{figure_HANNA}, we show a $T, x, y$-diagram for the heteroazeotropic, binary mixture butanol -- water generated by the CEM, based on activity coefficients from the HANNA-model (Antoine parameters, melting temperatures, and melting heats were taken from \citep{nist2024}). As can be seen, the whole range of phase equilibria, ranging from SLE, over LLE, over VLLE, to VLE is contained in the $T, x, y$-diagram. 

\subsection{Quantitative Evaluation}

To provide a quantitative evaluation of our framework, we use the CEM to calculate phase equilibria for various mixtures and compare our results to experimental data from the literature \citep{oh2003, kim2011, xiao2013, li2016, li2017, gao2018}. We measure the accuracy of the generalized CEM by MD to given data:
\begin{equation}
\textmd{MD}=\frac{\sum_{i=1}^{N}\sum_{j=1}^{M}\sum_{k=1}^{P}|x_{ijk}^{\text{source}}-x_{ijk}^{\text{CEM}}|}{NMP},
\end{equation}
where $N$ is the number of components, $M$ the number of examined feed streams, $P$ the number of phases and $x^{\text{source}}$ refers to the molar fraction reported in the literature. 

As in \citep{goettl2023}, we set the discretization parameter $\delta$ to $256$ for binary mixtures, to $128$ for ternary mixtures, and to $64$ for quaternary mixtures. If the CEM fails to calculate the correct number of phases or the correct aggregate states of the phases, we will note this data point as a missing feed (MF). These occur only rarely, for example, if the feed composition is very close to an azeotrope.

Tables \ref{Tab.binVLE.pxy}, \ref{Tab.binVLE.Txy}, and \ref{Tab.binSLE.Txy} show the results for binary VLEs and SLEs, respectively. The value of the mean deviation is below 0.015 for all mixtures, deducing a high accuracy of the calculated phase splits. Furthermore, there are no mising feeds for SLEs. The missing feeds for the VLEs occur exclusively close to the boundary of the two-phase regions, e.g., close to an azeotrope.

\begin{table}
\caption{Results for binary VLEs from \citep{oh2003} at 313.15 K ($p, x, y$-data). $M$ describes the number of feed streams, which were examined for the calculation of the mean deviation MD and MF is the number of missing feed streams.}
\label{Tab.binVLE.pxy}
\begin{center}
\begin{tabular}{|p{5.5cm}||c|c|c|c|}
    \hline
    Mixture & $p$ $\backslash$ bar & $M$ & MF & MD \\
    \hline
    methanol--benzene & 0.20--0.60 & 23 & 1 & 0.006 \\
    \hline
    methanol--toluene & 0.05--0.60 & 23 & 3 & 0.014 \\
    \hline
    MTBE--benzene & 0.20--0.70 & 23 & 2 & 0.006 \\
    \hline
    MTBE--methanol & 0.30--0.75 & 24 & 4 & 0.004 \\
    \hline
    MTBE--toluene & 298.15 & 20 & 0 & 0.005 \\
    \hline
\end{tabular}
\end{center}
\end{table}

\begin{table}
\caption{Results for binary VLE from \citep{xiao2013} at 1.013 bar ($T, x, y$-data). $M$ describes the number of feed streams, which were examined for the calculation of the mean deviation MD and MF is the number of missing feed streams.}
\label{Tab.binVLE.Txy} 
\begin{center}
\begin{tabular}{|p{5.5cm}||c|c|c|c|}
    \hline
    Mixture & $T$ $\backslash$ K & $M$ & MF & MD \\
    \hline
    methyl acetate--isopropyl acetate & 320.0--370.0 & 17 & 0 & 0.012 \\
    \hline
\end{tabular}
\end{center}
\end{table}

\begin{table}
\caption{Results for binary SLEs from \citep{kim2011} and \citep{gao2018} at 1.013 bar ($T, x, y$-data). $M$ describes the number of feed streams, which were examined for the calculation of the mean deviation MD and MF is the number of missing feed streams.}
\label{Tab.binSLE.Txy}
\begin{center}
\begin{tabular}{|p{5.5cm}||c|c|c|c|}
    \hline
    Mixture & $T$ $\backslash$ K & $M$ & MF & MD \\
    \hline
    chloroform -- acetylacetone & 180.0--260.0 & 13 & 0 & 0.002 \\
    \hline
    2-naphthaldehyde -- ethyl acetate & 283.15--313.15& 7 & 0 & 0.000\\
    \hline
    4-methylphthalic anhydride -- ethyl acetate & 283.15--313.15 & 7 & 0 &0.001\\
    \hline
\end{tabular}
\end{center}
\end{table}

The results for the ternary VLEs and SLEs are reported in Tables \ref{Tab.ternVLE.pxy} and \ref{Tab.ternSLE}. Again, there are no missing feeds in case of the SLEs. For the VLE mixtures, the single missing feed is close to the boundary of the multiphase region. The mean deviation is below 0.021 for all mixtures indicating that the CEM is also able to calculate ternary VLEs and SLEs with high accuracy. 

\begin{table}
\caption{Results for ternary VLEs from \citep{oh2003} at 313.15 K ($p, x, y$-data). $M$ describes the number of feed streams, which were examined for the calculation of the mean deviation MD and MF is the number of missing feed streams.}
\label{Tab.ternVLE.pxy}
\begin{center}
\begin{tabular}{|p{5.5cm}||c|c|c|c|}
    \hline
    Mixture & $p$ $\backslash$ bar & $M$ & MF & MD \\
    \hline
    MTBE--methanol--benzene & 0.30--0.75 & 67 & 1 & 0.010 \\
    \hline
    MTBE--methanol--toluene & 0.20--0.75 & 67 & 0 & 0.017 \\
    \hline
\end{tabular}
\end{center}
\end{table}

\begin{table}
\caption{Results for ternary SLEs from \citep{li2016}. $M$ describes the number of feed streams, which were examined for the calculation of the mean deviation MD and MF is the number of missing feed streams.}
\label{Tab.ternSLE}
\begin{center} 
\begin{tabular}{|p{5.5cm}||c|c|c|c|c|}
    \hline
    Mixture & $T$ $\backslash$ bar & $p$ $\backslash$ bar & $M$ & MF & MD \\
    \hline
    3,4-dichloronitrobenzene -- 2,3-dichloronitrobenzene -- ethanol & 283.15&1.0133& 11&0&0.002\\
    \hline
    3,4-dichloronitrobenzene -- 2,3-dichloronitrobenzene -- ethanol & 293.15&1.0133&15 &0&0.005\\
    \hline
    3,4-dichloronitrobenzene -- 2,3-dichloronitrobenzene -- ethanol & 303.15&1.0133&15 &0&0.007\\
    \hline
    3,4-dichloronitrobenzene -- 2,3-dichloronitrobenzene -- n-propanol & 283.15&1.0133&12 &0&0.002\\
    \hline
    3,4-dichloronitrobenzene -- 2,3-dichloronitrobenzene -- n-propanol & 293.15&1.0133&14 &0&0.006\\
    \hline
    3,4-dichloronitrobenzene -- 2,3-dichloronitrobenzene -- n-propanol & 303.15&1.0133&15 &0&0.020\\
    \hline
\end{tabular}
\end{center}
\end{table}

In Tables \ref{Tab.quatVLE.Txy} and \ref{Tab.quatSLE}, the results show that the CEM calculates quaternary VLEs and SLEs with high accuracy.

It should be noted that in the comparisons shown in Tables \ref{Tab.binVLE.pxy}-\ref{Tab.quatSLE}, there is an inherent deviation coming from the fact that the provided thermodynamic models deviate from the experimental data. Additional deviations come from the discretization procedure of the CEM. We only report the combined deviations.

\begin{table}
\caption{Results for quaternary VLE mixture from \citep{xiao2013} at 1.013 bar ($T, x, y$-data). $M$ describes the number of feed streams, which were examined for the calculation of the mean deviation MD and MF is the number of missing feed streams.}
\label{Tab.quatVLE.Txy}
\begin{center}
\begin{tabular}{|p{5.5cm}||c|c|c|c|}
    \hline
    Mixture & $T$ $\backslash$ K & $M$ & MF & MD \\
    \hline
    methyl acetate--methanol--isopropanol--isopropyl acetate & 320.0--360.0 & 16 & 0 & 0.022 \\
    \hline
\end{tabular}
\end{center}
\end{table}

\begin{table}
\caption{Results for quaternary SLEs from \citep{li2017}. $M$ describes the number of feed streams, which were examined for the calculation of the mean deviation MD and MF is the number of missing feed streams.}
\label{Tab.quatSLE}
\begin{center}
\begin{tabular}{|p{5.5cm}||c|c|c|c|c|}
    \hline
    Mixture & $T$ $\backslash$ bar & $p$ $\backslash$ bar & $M$ & MF & MD \\
    \hline
    adipic acid--glutaric acid--succinic acid--ethanol & 283.15&1.012&21 &2&0.001 \\
    \hline
    adipic acid--glutaric acid--succinic acid--ethanol & 303.15&1.012&24 &1&0.004 \\
    \hline
    adipic acid--glutaric acid--succinic acid--ethanol & 313.15&1.012&22 &0&0.008 \\
    \hline
\end{tabular}
\end{center}
\end{table}

\subsection{$g^E$-model parameter fitting using the CEM}
\label{section_ge_fit_results}

As explained beforehand, the CEM can be used to fit parameters of a $g^E$-model to given experimental data. In Table \ref{table_ge_fit}, we show results for carrying out the parameter fit for the NRTL model using the genetic algorithm explained in Section \ref{section_method_ge_fit}. We set the number of generations and the population size for the genetic algorithm both to $50$.

\begin{landscape}

\begin{table}
\caption{Results of parameter fit for binary VLEs and SLEs from \citep{kim2011}, \citep{oh2003}, \citep{xiao2013}. The tuple $(A_{12}, A_{21}, \alpha)$ refers to the found NRTL parameters ($A_{12}, A_{21}$ are defined as in Equation \ref{NRTL_para_def}), MD is the mean deviation and MF is the number of missing feed streams using those parameters for the calculation of the respective phase equilibria. The parameter $\alpha$ was constrained to the range $(0.1, 1)$.}
\label{table_ge_fit}
\begin{center}
\begin{tabular}{|p{5.cm}||c|c|c|c|c|c|c|}
\hline
Mixture & $T$ $\backslash$ K & $p$ $\backslash$ bar & $A_{12}$ & $A_{21}$ & $\alpha$ & MF & MD\\
\hline
chloroform--acetylacetone & 180.0--260.0 & 1.0133 & $9.6023\times10^2$& $-2.8627\times10^3$ & $2.6205\times10^{-1}$ & 0 & 0.002 \\
\hline
methanol--benzene & 313.15 & 0.2--0.6  & $1.6776\times10^3$ & $4.2855\times10^3$ & $3.2696\times10^{-1}$ & 0 & 0.010 \\
\hline
methanol--toluene & 313.15 & 0.05--0.6 & $3.348\times10^3$7 & $6.3418\times10^3$ & $4.8615\times10^{-1}$ & 0 & 0.028 \\
\hline
MTBE--benzene & 313.15 & 0.2--0.7  & $-5.3803\times10^3$ & $8.2421\times10^3$ & $1.4228\times10^{-1}$ & 0 & 0.012 \\
\hline
MTBE--methanol & 313.15 & 0.3--0.75 & $1.9560\times10^3$ & $9.1785\times10^2$ & $1.5389\times10^{-1}$  & 2 & 0.003 \\
\hline
MTBE--toluene & 313.15 & 0.05--0.7 & $2.7334\times10^3$ & $-7.1639\times10^2$ & $8.0500\times10^{-1}$  & 0 & 0.004 \\
\hline
methyl acetate--isopropyl acetate & 320.0--370.0 & 1.013 & $-4.6044\times10^3$ & $5.7841\times10^3$&  $1.0131\times10^{-1}$  & 0 & 0.009 \\
\hline
\end{tabular}
\end{center}
\end{table}

\end{landscape}

\section{Discussion}

The present work demonstrates a generalization of the CEM \citep{ryll2009, ryll2012, goettl2023} that allows to do $T, p$ flash calculations for mixtures involving solid, liquid, and gaseous aggregate states. This approach enables the determination of arbitrary phase equilibria through a systematic discretization of the entire composition space and the subsequent construction of the convex envelope for the Gibbs energy. Building upon the mathematical framework proposed by \citep{goettl2023}, the method is designed to handle mixtures with an arbitrary number of components and phases. The CEM can also be used indirectly to qualitatively identify points or regions where phase behavior changes, such as variations in the number and type of phases, for example at (hetero-)azeotropes. As shown in Figure 4, one can iterate over a range of temperatures and monitor the phase behavior to identify transitions between vapor and liquid phases (in Figure 4, a heteroazeotrope and a solid–liquid-liquid equilibrium are observed).

Quantitative validation was achieved for VLEs and SLEs involving up to four components (LLEs were already addressed in \citep{goettl2023}). Regardless of the number of components, the VLEs and SLEs could be calculated reliably and with a high degree of accuracy. In rare cases, and only near the boundaries of multiphase regions, such as close to azeotropes, the CEM may occasionally struggle to predict the correct phase splits. These limitations arise from two main sources. First, the CEM operates on a discretized composition space, meaning that phase boundaries are approximated based on a finite set of evaluated points. While increasing the resolution of the discretization can improve accuracy, it cannot completely eliminate interpolation errors. Moreover, a finer discretization comes at the cost of increased computational effort during the convex envelope construction. Second, the thermodynamic models used in the CEM are taken as given and have been typically fitted to experimental data beforehand. As a result, they inherently carry modeling inaccuracies and reflect the uncertainty or noise present in the underlying data. These factors limit the quantitative precision of the computed phase behavior, particularly close to the boundaries of multiphase regions. Nonetheless, the deterministic nature of the convex envelope construction ensures that the method always yields a consistent result, and overall reliability remains high across the examined systems.

Additionally, it is shown how to use the CEM for fitting parameters of $g^E$-models, such as NRTL, using experimental data. The methodology, detailed in Section \ref{section_method_ge_fit}, can also be applied to mixtures with more than two components, although computational complexity poses a limitation. For instance, a $T, p$ flash over the entire composition space for a ternary mixture requires approximately one second on a AMD Ryzen 5 5500U CPU. However, since experimental data often spans varying $T$ and $p$, completing a full cycle over the dataset becomes time-intensive. This challenge is further amplified by the computational demands of the utilized genetic algorithm, which requires numerous full cycles to converge (e.g., $50\times50=2500$ for the shown results from Section \ref{section_ge_fit_results}).

While in theory (as proven mathematically in \citep{goettl2023}), the CEM can be applied to mixtures of an arbitrary number of components, computational complexity becomes also a limiting factor, when mixtures with more than six components are examined. Particularly the construction of the convex envelope is a challenging task. This also applies for the generalized framework presented in this work and therefore, we refer to \citep{goettl2023} for a thorough discussion of the implications of that matter.

Clearly, the discretization resolution impacts the accuracy of the calculation results. The resolution to be chosen depends on the application and on how accurately the composition of the phases in equilibrium needs to be known. Additionally, a fine-grained discretization increases the computational effort required to construct the convex hull and also raises the number of simplices that need to be classified. Therefore, this usually imposes a constraint on the discretization for mixtures with a larger number of components. An interesting direction for future research could be to investigate strategies for adaptive discretization for the CEM.

Another potential direction for future research lies in incorporating more advanced models for calculating the Gibbs energy of mixing. For example, more complex solid-phase behavior could be incorporated, in contrast to the present work, where only pure components are allowed to form solid phases. Additionally, the framework could be expanded to include the chemical equilibria of reactive mixtures, a concept previously described and implemented for binary and ternary mixtures by \citep{hildebrandt1994}.

As demonstrated, the CEM can be effectively combined with predictive tools such as the HANNA model \citep{specht2024} to perform qualitative analyses of mixtures and their phase equilibria. This approach could be further developed by integrating additional models that predict pure component properties, such as melting temperature, melting enthalpy, or vapor pressure, as proposed in \citep{lansford2023} and \citep{santana2024}. Moreover, future research could explore combining CEM-based simulations with conceptual process evaluation methods, as previously shown in \citep{ryll2009, goettl2021, goettl2024}. Such an integration could unify approaches for solvent and process design by predicting thermodynamic properties from SMILES-strings, simulating the resulting phase equilibria, and modeling apparatuses, e.g., for decantation or distillation, using the CEM.

\section{Conclusion}

This work generalizes the CEM to encompass mixtures with solid, liquid, and gaseous aggregate states, enabling the determination of arbitrary phase equilibria without prior knowledge about the number of occurring phases. By leveraging a robust mathematical framework, the methodology is shown to be applicable to multi-component mixtures. It has been quantitatively validated for VLEs and SLEs involving up to four components, demonstrating high accuracy of the phase splits. Additionally, the present work demonstrates how the CEM can be utilized to fit parameters for $g^E$ models using experimental data. In summary, the CEM offers a robust method for $T, p$ flash calculations, without requiring any \textit{a priori} knowledge about the phase splits, e.g., the number of phases or respective aggregate states. Looking ahead, the combination of the CEM with a machine learning model for activity coefficient prediction illustrates its potential for integration into simulations, paving the way for unifying solvent and process design in future research.

\bibliographystyle{elsarticle-harv} 
\bibliography{ref}





\end{document}